\documentclass{article}
\usepackage{spconf,amsmath,graphicx}
\usepackage{multirow}
\usepackage{amsfonts}
\usepackage{mathtools}
\usepackage{url}

\title{On Multiangle Discrete Fractional Periodic Transforms}

\name{Christian Oswald, Franz Pernkopf
\thanks{Research funded by the Austrian Research Promotion Agency (FFG), Infineon Technologies Austria AG and Graz University of Technology under the REPAIR project (40352729)}
\thanks{© 2026 IEEE. Personal use of this material is permitted. Permission from IEEE must be obtained for all other uses, in any current or future media, including reprinting/republishing this material for advertising or promotional purposes, creating new collective works, for resale or redistribution to servers or lists, or reuse of any copyrighted component of this work in other works.}
\thanks{This work has been published at 2026 IEEE International Conference on Acoustics, Speech and Signal Processing (ICASSP 2026). DOI: 10.1109/ICASSP55912.2026.11460624}}
\address{Graz University of Technology, Austria}

\begin{document}

\ninept
\maketitle
\begin{abstract}
The efficient multiangle centered discrete fractional Fourier transform (MA-CDFRFT) \cite{vargas2005multiangle} has proven to be a useful tool for time-frequency analysis; in this paper, we generalize the MA-CDFRFT to general $M$-periodic transforms, which, among others, include the \textit{standard} discrete Fourier,
discrete sine, discrete cosine, Hadamard and discrete Hartley transform. Furthermore, we exploit the symmetries inherent to the MA-CDFRFT and our novel multiangle standard discrete fractional Fourier transform (MA-DFRFT) to halve the number of FFTs needed to compute these transforms, which paves the way for applications in resource-constrained environments. 
\end{abstract}
\section{Introduction} \label{sec:introduction}
Fractional transforms such as the discrete fractional Fourier transform (DFRFT) are important tools in digital signal processing, as they are used for linearly frequency modulated (LFM) chirp estimation, compression and mitigation \cite{wang2012sar, oswald2026fmcw, oswald2025fractional}, cryptography \cite{rawat2012blind}, improved spectrograms \cite{vargas2005improved}, processing optical systems \cite{torre2002fractional}, among other applications \cite{sejdic2011fractional}. 
Many of these applications require computing multiple transforms of different fractional orders; for instance, \cite{oswald2026fmcw, oswald2025fractional} does this to check for the presence of LFM chirp interferences in radar signals, as DFRFTs of different fractional orders pulse-compress interferences with different chirp rates.
In Sec. \ref{sec:applications}, we discuss some applications in more detail.

The multiangle centered discrete fractional Fourier transform (MA-CDFRFT) \cite{vargas2005multiangle} utilizes FFTs to efficiently compute the \mbox{CDFRFT} in parallel for multiple fractional orders -- also commonly referred to as fractional \textit{angles} in the context of the fractional Fourier transform.
In this paper, we generalize the approach in \cite{vargas2005multiangle} to all $M$-periodic transforms. These include the \textit{standard} DFT as their most prominent member, which, compared to the \textit{centered} DFT discussed in \cite{vargas2005multiangle}, is arguably better explored and more relevant for practical applications. We call our extension of the MA-CDFRFT to the standard DFT the multiangle discrete fractional Fourier transform (MA-DFRFT).
In the quest of implementing the MA-(C)DFRFT as efficiently as possible for applications in resource-constrained environments, we furthermore leverage the symmetries present in the MA-(C)DFRFT to halve the number of FFTs required to compute these transforms\footnote{We provide Python code at \url{https://github.com/OsChri}.}. 

We denote vectors and sets as lowercase, and matrices as uppercase boldface letters. Their elements are indexed using square brackets; for instance, $\boldsymbol{A}[n,k]$ references the entry of $\boldsymbol{A}$ in row $n$ and column $k$. To denote the entire $n^\mathrm{th}$ row of $\boldsymbol{A}$, we write $\boldsymbol{A}[n, :]$, while for the entire $k^\mathrm{th}$ column, we use \mbox{$\boldsymbol{A}[:,k]$}. We sometimes call the DFT the \textit{standard} DFT for better distinction from the \textit{centered} DFT (CDFT); the same applies to their fractional generalizations, the standard DFRFT and the CDFRFT, which we jointly address as the (C)DFRFT.
We use subscripts to refer to specific instances of vectors and matrices, e.g., we use $\{\cdot\}_c$ and $\{\cdot\}_s$ to distinguish between variables for the centered and the standard DFT, respectively. 
If a relation holds for all $M$-periodic matrices, we omit these subscripts; for example, $\boldsymbol{W}_s$ denotes the DFT matrix, while $\boldsymbol{W}$ is a general $M$-periodic matrix, which, when thought of as a set, contains $\boldsymbol{W}_s$. In this paper, we use the terms \textit{matrix} and \textit{transform} interchangeably. 

\section{Preliminaries} \label{sec:background}
\subsection{$M$-periodic matrices} \label{sec:m_periodic}
Let $\boldsymbol{W} \in \mathbb{C}^{N \times N}$ be an $M$-periodic matrix, i.e., $\boldsymbol{W}^M = \boldsymbol{I}$, $M \in \mathbb{N}$, where $\boldsymbol{I}$ is the identity matrix. $\boldsymbol{W}$ is $M$-periodic if and only if
\begin{enumerate}
    \item it is diagonizable over $\mathbb{C}$ and
    \item all its eigenvalues are the $M^\text{th}$ roots of unity.
\end{enumerate}
We write the eigenvalues of $\boldsymbol{W}$ as a diagonal matrix 
\begin{equation*}
    \boldsymbol{\Lambda} = \text{diag}(\omega_M^{\boldsymbol{l}}),
\end{equation*}
where $\omega_M = e^{-j \frac{2\pi}{M}}$ is a primitive $M^{\textrm{th}}$ root of unity and $\boldsymbol{l} \in \mathbb{N}_0^N$. 

We can compute fractional powers $\boldsymbol{W}^a$, where $a \in \mathbb{R}$ is the so-called fractional order through its eigendecomposition $\boldsymbol{W}^a = \boldsymbol{V} \boldsymbol{\Lambda}^a \boldsymbol{V}^{-1}$, where $\boldsymbol{V}$ are the eigenvectors of $\boldsymbol{W}$. Noninteger powers of the complex exponentials within $\boldsymbol{\Lambda}$ are multivalued, but if we pick a certain branch by defining $(e^{j\theta})^a \coloneq e^{j\theta a}$, $\forall \ \theta \in \mathbb{R}$, we get $\boldsymbol{W}^{a_1} \boldsymbol{W}^{a_2} = \boldsymbol{W}^{a_2} \boldsymbol{W}^{a_1} = \boldsymbol{W}^{{a_1}+{a_2}}, \ \forall \ {a_1},{a_2} \in \mathbb{R}$, which is commonly known as \textit{angle-additivity}. 

Many important transforms are instances of $\boldsymbol{W}$, such as the (C)DFT ($M=4$), the type-I and type-IV discrete sine (DST) and cosine (DCT) transforms ($M=2$), the Hadamard transform (HT) ($M=2$), the discrete Hartley transform (DHT) ($M=2$) \cite{pei2001discrete, pei1999discretehadamard, tseng2002eigenvalues}, all Householder reflections ($M=2$), all permutation matrices including the cyclic shift $\boldsymbol{S}$ ($M=N$), where
\begin{equation*}
    \boldsymbol{S}[n,m] = 
    \begin{cases}
    1 & \text{if } n = (m + 1) \text{ mod } N, \\  
    0 & \text{otherwise},
\end{cases}
\end{equation*}
among others. The eigenvalue multiplicities of some commonly used $M$-periodic transforms are summarized in Tab. \ref{tab:ev_dft} and Tab. \ref{tab:ev_m2} as a function of $N$. Following their eigenvalue multiplicities we can write \mbox{$\boldsymbol{l} = \boldsymbol{l}_{cso} \coloneq ( 0, 1, 2,\ldots, N-1)^T$} for the \underline{C}DFT, the \underline{s}tandard \underline{o}dd-length DFT, the type-I and type-IV DST and DCT, the HT, the DHT when $N \text{ mod } 4 \neq 0$ as well as $\boldsymbol{S}$. Meanwhile, $\boldsymbol{l} = \boldsymbol{l}_{se} \coloneq (0, 1, 2, \ldots, N-3, N-2,N)^T$ for the \underline{s}tandard \underline{e}ven-length DFT and the DHT with $N \text{ mod } 4 = 0$. 
\begin{table}[t]
\caption{(C)DFT and DHT eigenvalue multiplicities}
\begin{center}
\begin{tabular}{c|c||c|c|c|c}
& $N $& $1$ & $-j$ & $-1$ & $j$ \\
\hline
\hline
\multirow{4}{*}{DFT} &$4m$ & $m+1$ & $m$ & $m$ & $m-1$ \\
& $4m+1$ & $m+1$ & $m$ & $m$ & $m$ \\
& $4m+2$ & $m+1$ & $m$ & $m+1$ & $m$ \\
& $4m+3$ & $m+1$ & $m+1$ & $m+1$ & $m$ \\
\hline
\hline
\multirow{4}{*}{CDFT} & $4m$ & $m$ & $m$ & $m$ & $m$ \\
& $4m+1$ & $m+1$ & $m$ & $m$ & $m$ \\
& $4m+2$ & $m+1$ & $m+1$ & $m$ & $m$ \\
& $4m+3$ & $m+1$ & $m+1$ & $m+1$ & $m$ \\
\hline
\hline
\multirow{4}{*}{DHT} & $4m$ & $2m + 1$ & $0$ & $2m - 1$ & $0$ \\
& $4m+1$ & $2m+1$ & $0$ & $2m$ & $0$ \\
& $4m+2$ & $2m+1$ & $0$ & $2m+1$ & $0$\\
& $4m+3$ & $2m+2$ & $0$ & $2m+1$ & $0$ 
\end{tabular}
\label{tab:ev_dft}
\end{center}
\end{table}

\begin{table}
\begin{center}
\caption{HT, DST-I, DST-IV, DCT-I and DCT-IV eigenvalue multiplicities}
\begin{tabular}{c||c|c}
$N$ & $1$ & $-1$ \\
\hline
\hline
$2m$ & $m$ & $m$ \\
$2m+1$ & $m+1$ & $m$
\end{tabular}
\label{tab:ev_m2}
\end{center}
\end{table}

The choice of $\boldsymbol{l}$ is not unique for a given $\boldsymbol{W}$; every $\boldsymbol{l}$ that reduces to the same $\boldsymbol{l}_w = \boldsymbol{l} \text{ mod } M$ (in an element-wise manner) corresponds to the same eigenvalue multiplicities, where $\boldsymbol{l}_w$ simply keeps the argument of the eigenvalues within $(-2 \pi, 0]$. 
For $z \in \mathbb{Z}$, any $\boldsymbol{l}$ corresponding to the same eigenvalue multiplicities lead to the same $\boldsymbol{W}^z$, while for $p \in \mathbb{R} \backslash \mathbb{Z}$, $\boldsymbol{W}^p$ is different for every $\boldsymbol{l}$; $\boldsymbol{l}_{cso}$ and $\boldsymbol{l}_{se}$ follow the definition of the \textit{chirp fractional Fourier transform} \cite{cariolaro2002multiplicity}, which can be used for LFM chirp compression when choosing Hermite-Gauss-like eigenvectors \cite{candan2000discrete}. Meanwhile, $\boldsymbol{l}_{csow} = \boldsymbol{l}_{cso} \text{ mod } M$ and $\boldsymbol{l}_{sew} = \boldsymbol{l}_{se} \text{ mod } M$ correspond to the \textit{\underline{w}eighted fractional Fourier transform} \cite{cariolaro2002multiplicity}, which does not compress LFM chirps. 

\subsection{DFT, CDFT, DFRFT and CDFRFT} \label{sec:dft_cdft}
The unitary standard DFT matrix $\boldsymbol{W}_s$ of size $N \times N$  is defined as 
\begin{equation}
    \boldsymbol{W}_s[n,k] = \frac{1}{\sqrt{N}} \omega_N^{nk}.
\end{equation}
In contrast, the $N \times N$ unitary centered DFT matrix $\boldsymbol{W}_c$ is given by
\begin{equation}
    \boldsymbol{W}_c[n,k] = \frac{1}{\sqrt{N}} \omega_N^{(n-\frac{N-1}{2})(k-\frac{N-1}{2})}.
\end{equation}
For odd $N$, $\boldsymbol{W}_s$ and $\boldsymbol{W}_c$ are identical except for a row- and column-wise circular shift of their elements. However, for even $N$, the standard and the centered DFT probe different frequency components of their input. More specifically, the CDFT evaluates frequencies which fall exactly in between the standard DFT frequencies; for instance, if $N$ even, the CDFT does not compute a signal's zero-frequency component. 

As already mentioned in Sec. \ref{sec:m_periodic}, $\boldsymbol{W}_c$ and $\boldsymbol{W}_s$ are instances of $\boldsymbol{W}$ with $M = 4$. For powers $4z,4z+1,4z+2,4z+3$, $z \in \mathbb{Z}$, the (C)DFT becomes the identity, the forward (C)DFT, a reversal matrix $\boldsymbol{P}$ and the inverse (C)DFT, respectively. For the CDFT and the odd-length DFT, $\boldsymbol{P} = \boldsymbol{P}_{cso}$, which has ones on the anti-diagonal and zeros otherwise, while for the even-length DFT, $\boldsymbol{P} = \boldsymbol{P}_{se}$, where
\begin{equation}
\boldsymbol{P}_{se}[n,k] = 
\begin{cases}
1, & \text{if } n=k=0, \\
1, & \text{if } n+k=N, \\ 
0, & \text{otherwise.}
\end{cases} \label{eq:p_s}
\end{equation}
Due to the (C)DFT's eigenvalue multiplicities listed in Tab. \ref{tab:ev_dft}, its eigenvectors are not unique when $N \geq 4$, which led to numerous proposals for possible favorable choices. Several publications \cite{candan2000discrete, serbes2011discrete, de2017discrete}, among others, construct standard DFT eigenvectors that approximate the continuous Fourier transform's eigenfunctions, which are the Hermite-Gaussian functions. (C)DFRFT realizations built with such eigenvectors (whilst choosing $\boldsymbol{l}_{cso}$ or $\boldsymbol{l}_{se}$ following the definition of the chirp fractional Fourier transform \cite{cariolaro2002multiplicity}) are useful for LFM chirp processing, as then a (C)DFRFT of order $a$ compresses an LFM chirp with chirp rate $\textrm{arctan}(a \pi/2)$ into a small number of samples; this mimics the continuous fractional Fourier transform of order $a$, which rotates its input's Wigner-Ville distribution by $a \pi/2$ radians \cite{almeida1993introduction}. A comparison of CDFRFT realizations in terms of their chirp processing capabilities can be found in \cite{peacock2013comparison, bhatta2015comparative}.
Other eigenvectors of the standard DFT are sparse and have repeating elements \cite{de2019reduced}, a high degree of symmetry \cite{erseghe2003orthonormal} or are generated by a complete generalized Legendre sequence \cite{pei2008closed}; a survey can be found in \cite{su2019analysis}. However, a "universally optimal" choice for the (C)DFT eigenvectors remains a subject of current research.

\section{Efficient Multiangle Fractional $M-$periodic Transforms} \label{sec:ma_ft_simple}
In \cite{vargas2005multiangle}, the authors have used the eigenvalue multiplicities of the CDFT to construct an algorithm which efficiently computes the multiangle CDFRFT (MA-CDFRFT), i.e., the CDFRFTs of fractional orders $\{0, 4/N, \dots, 4(N-1)/N\}$ in parallel. A naive approach has complexity $\mathcal{O}(N^3)$ due to the $N$ matrix-vector multiplications involved, while they use FFTs to reduce this complexity to $\mathcal{O}(N^2 \log N)$. In this section, we generalize their method so that it applies to any $M$-periodic transform $\boldsymbol{W}$. 

First, we expand and reorder $\boldsymbol{W}^a\boldsymbol{x}$, $\boldsymbol{x} \in \mathbb{C}^N$ being the input signal, that is, 
\begin{align}
   (\boldsymbol{W}^a\boldsymbol{x})[n] &= \sum_{i=0}^{N-1} \boldsymbol{x}[i] \boldsymbol{W}^a[n,i] \nonumber\\
    &= \sum_{i=0}^{N-1} \boldsymbol{x}[i] \sum_{k=0}^{N-1} \boldsymbol{V}[n,k] \boldsymbol{V}^{-1}[k,i] \omega_M^{a\boldsymbol{l}[k]} \nonumber\\
    &= \sum_{k=0}^{N-1} \sum_{i=0}^{N-1} \boldsymbol{x}[i]  \boldsymbol{V}[n,k] \boldsymbol{V}^{-1}[k,i] \omega_M^{a\boldsymbol{l}[k]} \nonumber\\
     &= \sum_{k=0}^{N-1} \boldsymbol{Z}[k,n] \omega_{M}^{a\boldsymbol{l}[k]},\nonumber \\
    \boldsymbol{Z}[k,n] 
    &= \boldsymbol{V}[n,k]  (\boldsymbol{V}^{-1} \boldsymbol{x})[k]. \label{eq:pre_fft_c}
\end{align}
Now we replace $a$ with a vector of special fractional orders $\boldsymbol{a}_M[r] =  rM/N$, $r \in \{0, 1,2, \ldots, N-1\}$ to define the multiangle fractional transform 
\begin{equation}
    \boldsymbol{X}[r,n] \coloneq \sum_{k=0}^{N-1}\boldsymbol{Z}[k,n] \omega_M^{rM\boldsymbol{l}[k]/N} = \sum_{k=0}^{N-1}\boldsymbol{Z}[k,n] \omega_N^{r\boldsymbol{l}[k]}, \label{eq:ma_cdfrft}
\end{equation}
i.e., $\boldsymbol{X}[r,:] = (\boldsymbol{W}^{rM/N}\boldsymbol{x})^T$, where $\{\cdot\}^T$ indicates transposition. 
Rewriting \eqref{eq:ma_cdfrft}, we find that 
\begin{equation*}
    \boldsymbol{X} = \boldsymbol{B}\boldsymbol{Z},
\end{equation*}
where $\boldsymbol{B}[r,k] = \omega_N^{r\boldsymbol{l}[k]}$. 

We now exploit the special structure of $\boldsymbol{B}$ to reduce the computational complexity of calculating $\boldsymbol{B} \boldsymbol{Z}$ from $\mathcal{O}(N^3)$ to $\mathcal{O}(N^2 \log N)$. More concretely, we factorize \mbox{$\boldsymbol{B} = \sqrt{N} \boldsymbol{W}_s \boldsymbol{\Gamma}$} to efficiently compute $\boldsymbol{B}\boldsymbol{Z}$ through column-wise FFTs $\sqrt{N} \boldsymbol{W}_s$ and an additional transform $\boldsymbol{\Gamma}$.
\begin{figure*}[!t]
\centering
\includegraphics[width=\textwidth]{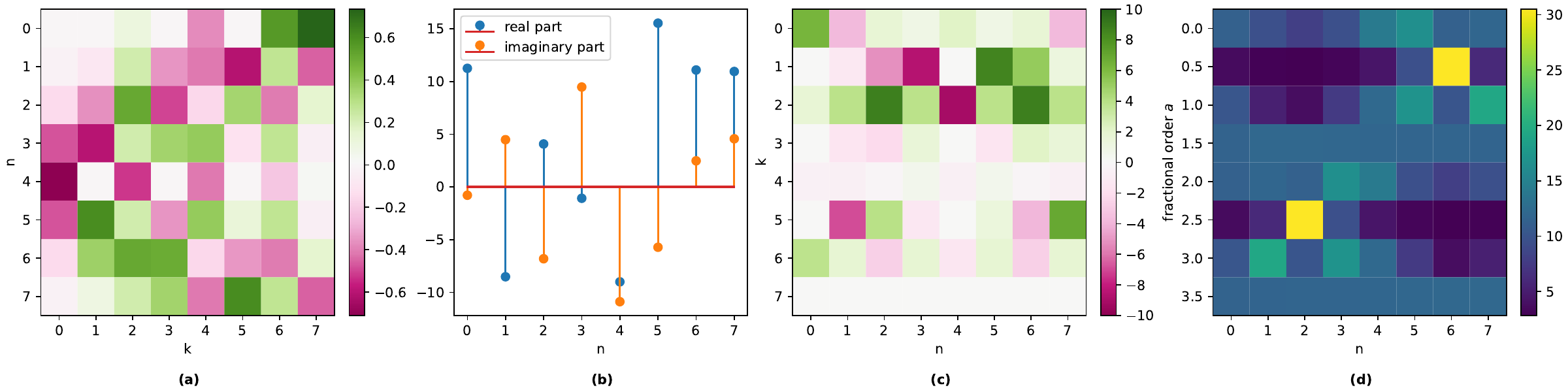}
\caption{\textbf{(a)} Eigenvectors $\boldsymbol{V}_{se}$ for $N=8$ generated with the method in \cite{candan2000discrete}. \textbf{(b)} some complex-valued signal $\boldsymbol{x}$ to demonstrate the MA-DFRFT. \textbf{(c)} corresponding real parts of $\hat{\boldsymbol{Z}}_{se}$ using the eigenvectors from \textbf{(a)}. \textbf{(d)} resulting MA-DFRFT magnitudes $|\boldsymbol{X}_{se}|$, which reveal the LFM chirp component within $\boldsymbol{x}$ with a chirp rate of 1 corresponding to fractional orders $0.5$ and $2.5$. Note the symmetries as described in \eqref{eq:v_s_even}, \eqref{eq:symmetries_n_even} and \eqref{eq:X_s_even}, respectively. \label{fig:mdfrft}}  
\end{figure*}
It follows that
\begin{equation*}
    \boldsymbol{\Gamma} = \frac{1}{\sqrt{N}}\boldsymbol{W}_s^{-1}\boldsymbol{B},
\end{equation*} 
that is, the columns of $\boldsymbol{\Gamma}$ consist of the inverse DFTs of the columns of $\boldsymbol{B}$. Therefore, 
\begin{equation*} 
\boldsymbol{\Gamma}[n, k] = 
\begin{cases}
    1 & \text{if } n = \boldsymbol{l}[k] \text{ mod } N, \\  
    0 & \text{otherwise}.
\end{cases}
\end{equation*}
Since $\boldsymbol{\Gamma}$ has exactly one nonzero entry per column, the matrix-vector product $\hat{\boldsymbol{Z}} =\boldsymbol{\Gamma}\boldsymbol{Z}[:,n]$ consists of at most $N$ additions; therefore, $\boldsymbol{B}\boldsymbol{Z}[:,n]$ is $\mathcal{O}(N \log N)$, which means that $\boldsymbol{B}\boldsymbol{Z}$ has complexity $\mathcal{O}(N^2 \log N)$. 
$\boldsymbol{BZ}$ is in fact the most expensive operation when computing $\boldsymbol{X}$ as \eqref{eq:pre_fft_c} has complexity $\mathcal{O}(N^2)$, resulting in an overall computational complexity of $\mathcal{O}(N^2 \log N)$ for all definitions \cite{cariolaro2002multiplicity} of all $M$-periodic transforms $\boldsymbol{W}$. 

Note that if $\boldsymbol{l} = \boldsymbol{l}_{cso}$, $\boldsymbol{\Gamma}$ becomes the identity, which means that the columns of $\boldsymbol{X}$ and $\boldsymbol{Z}$ relate by a simple FFT.
As mentioned in Sec. \ref{sec:m_periodic}, this is the case for many commonly used transforms, including the chirp fractional definition of the CDFT (as was shown in \cite{vargas2005multiangle}), the type-I and type-IV DST and DCT, the HT, the DFT for odd $N$ and the DHT for $N \text{ mod } 4 \neq 0$ as well as $\boldsymbol{S}$. 

For $\boldsymbol{S}$, we can easily verify the equivalence of the efficient multiangle transform to $N$ distinct fractional transforms with angles $\boldsymbol{a}_N[r]$. Since $\boldsymbol{S}$ is circulant, its eigenvectors are the DFT basis, i.e., $\boldsymbol{V_S} = \boldsymbol{W}_s$; in \eqref{eq:pre_fft_c}, we therefore compute an inverse DFT $\boldsymbol{W}_s^{-1} \boldsymbol{x}$ before modulating $N$ copies of $\boldsymbol{W}_s^{-1} \boldsymbol{x}$ with the sinusoids that make up $\boldsymbol{W}_s$. In \eqref{eq:ma_cdfrft} we then compute FFTs of these modulated signals, which, according to the DFT's frequency shift theorem, correspond to cyclically shifted versions of $\boldsymbol{x}$. 
This makes sense as $\boldsymbol{a}_N[r] = r$, i.e., $\boldsymbol{X_S}$ is a circulant matrix formed from $\boldsymbol{x}$. 

We can now derive the even-length standard MA-DFRFT and the multiangle discrete fractional Hartley transform of length $N \text{ mod }4 = 0$ with eigenvalue powers $\boldsymbol{l}_{se}$; even-length transforms are especially relevant in practice due to the efficiency of even-length FFTs. We find that
\begin{equation}
\hat{\boldsymbol{Z}}_{se}[k,n] = 
\begin{cases}
\boldsymbol{Z}_{se}[0,n] + \boldsymbol{Z}_{se}[N-1,n], & {\text{if } k=0}, \\
0, & \text{if } k=N-1, \\ 
\boldsymbol{Z}_{se}[k,n], & {\text{otherwise,}}
\end{cases} \label{eq:zs_hat}
\end{equation}
i.e., $\hat{\boldsymbol{Z}}_{se}$ is a simple modification of $\boldsymbol{Z}_{se}$ consisting of a single addition and deletion per column of $\boldsymbol{Z}_{se}$. 
An example of $\hat{\boldsymbol{Z}}_{se}$ and a MA-DFRFT $\boldsymbol{X}_{se}$ can be seen in Fig. \ref{fig:mdfrft}c and Fig. \ref{fig:mdfrft}d, respectively. 

\section{Halving the computational complexity of the MA-(C)DFRFT} \label{sec:halving}
In this section, we utilize the symmetries within the (C)DFT eigenvectors to halve the number of FFTs needed to compute the chirp fractional definition of the MA-(C)DFRFT. Note that these symmetries are present in all admissible eigenvector sets of the (C)DFT. We leave optimizations of other multiangle fractional transforms for future research. 

The CDFT and the odd-length standard DFT eigenvectors $\boldsymbol{V}_{cso}$ in any order consistent with $\boldsymbol{\Lambda}_{cso} = \text{diag}(\omega_4^{\boldsymbol{l}_{cso}})$ have the symmetry
\begin{equation}    
    \boldsymbol{V}_{cso}[N-n-1,k] = (-1)^k \boldsymbol{V}_{cso}[n,k], \label{eq:v_c_even}
\end{equation}
$n,k \in \{0,1,...,N-1 \}$, that is, they are alternatingly even and odd symmetric \cite{vargas2005multiangle, mcclellan1972eigenvalue}.
Note that in \eqref{eq:pre_fft_c}, the row $\boldsymbol{Z}[k, :]$ is formed by multiplying the $k$\textsuperscript{th} eigenvector $\boldsymbol{V}[:, k]$ with the scalar $( \boldsymbol{V}^{-1} \boldsymbol{x})[k]$; it follows that the symmetry property \eqref{eq:v_c_even} also holds for $\boldsymbol{Z}^T_{cso}$, i.e.,
\begin{equation}
    \boldsymbol{Z}_{cso}[k, N-n-1] = (-1)^k \boldsymbol{Z}_{cso}[k,n]. \label{eq:Z_cso}
\end{equation}

In contrast, the symmetries of the even-length standard DFT eigenvectors $\boldsymbol{V}_{se}$ in any order consistent with \mbox{$\boldsymbol{\Lambda}_{se} = \text{diag}(\omega_4^{\boldsymbol{l}_{se}})$} are 
\begin{equation}
\boldsymbol{V}_{se}[N-u,k] = ((-1)^k + 2 \delta [N-k-1]) \boldsymbol{V}_{se}[u,k], \label{eq:v_s_even}
\end{equation}
$u \in \{1,2,...,N-1 \}$, and $\delta[n]$ is $1$ if $n=0$ and $0$ otherwise; in other words, the eigenvectors corresponding to eigenvalues $\pm1$ are even symmetric, while eigenvectors corresponding to $\pm j$ are odd symmetric \cite{mcclellan1972eigenvalue}. An example for $\boldsymbol{V}_{se}$ can be seen in Fig. \ref{fig:mdfrft}a. 
Transitively, \eqref{eq:v_s_even} also holds for $\boldsymbol{Z}^T_{se}$, which means that the symmetries of $\hat{\boldsymbol{Z}}_{se}$ are
\begin{equation}
\hat{\boldsymbol{Z}}_{se}[k, N-u] = (-1)^k \hat{\boldsymbol{Z}}_{se}[k,u]\label{eq:symmetries_n_even}.
\end{equation}
$\hat{\boldsymbol{Z}}_{se}[0,u]$ is a sum of two even vectors and therefore remains even, while $\hat{\boldsymbol{Z}}_{se}[N-1,u] = 0$. An example of $\hat{\boldsymbol{Z}}_{se}$ is visualized in Fig. \ref{fig:mdfrft}c. 
\subsection{Even-Length Case}
Let us first consider the subset of $\boldsymbol{V}_{cso}$ for even $N$, i.e., the even-length CDFT eigenvectors $\boldsymbol{V}_{ce}$. We find that $\boldsymbol{X}_{ce}$, which consists of column-wise FFTs of $\boldsymbol{Z}_{ce}$ since $\boldsymbol{\Gamma}_{cso} = \boldsymbol{I}$, can be computed more efficiently by using the DFT's frequency shift relation on the symmetries of $\boldsymbol{Z}_{ce}$ \eqref{eq:Z_cso}; more concretely, \mbox{$\boldsymbol{X}_{ce}[:, N-n-1]$} can be obtained from \mbox{$\boldsymbol{X}_{ce}[:,n]$} through
\begin{equation}
\boldsymbol{X}_{ce}[\left(r +  N/2 \right) \ \mathrm{mod} \ N, N-n-1] = \boldsymbol{X}_{ce}[r,n].
 \label{eq:S_c_even}
\end{equation}
This means that we can acquire half of the even-length MA-CDFRFT by mirroring and circular shifting its complementary half, reducing the number of required FFTs from $N$ as described in \cite{vargas2005multiangle} to $N/2$.
This result is not surprising, as for even $N$, the fractional orders $\boldsymbol{a}_4$ evaluated by the MA-(C)DFRFT are such that 
\begin{equation}
    \boldsymbol{a}_4[r] =(\boldsymbol{a}_4[r+N/2] +2) \text{ mod } 4. \label{eq:even_n}
\end{equation}
Due to angle-additivity we know that
$\boldsymbol{W}^{a+2} = \boldsymbol{W}^a \boldsymbol{W}^2$ while $\boldsymbol{W}^2 = \boldsymbol{P}$ for the (C)DFT, i.e., rows $0$ through $N/2 -1$ of the even-length MA-(C)DFRFT are reversed versions of rows $N/2$ through $N-1$.

In analogy to the even-length MA-CDFRFT, the column-wise FFTs of $\hat{\boldsymbol{Z}}_{se}$ relate via
\begin{align}
\begin{split}
\boldsymbol{X}_{se}[\left(r +  N/2 \right) \ \mathrm{mod} \ N, N-u] &= \boldsymbol{X}_{se}\left[r, u \right],  \label{eq:X_s_even} 
\end{split}
\end{align}
this time using the DFT's frequency-shift property on \eqref{eq:symmetries_n_even}. An example of $\boldsymbol{X}_{se}$ can be seen in Fig. \ref{fig:mdfrft}d. This means that we can compute the entire even-length MA-DFRFT by only calculating the FFTs of, e.g., columns $0$ through $N/2$, and retrieving the remaining columns with \eqref{eq:X_s_even}. This result is consistent with $\boldsymbol{W}_{se}^2 =\boldsymbol{P}_{se}$. 

Note that thanks to \eqref{eq:S_c_even} and \eqref{eq:X_s_even}, we only need to compute and store half of $\boldsymbol{Z}$ when evaluating the MA-(C)DFRFT. 
For real-valued $\boldsymbol{x}$, the MA-(C)DFRFT additionally becomes symmetric about fractional orders $0$ and $2$ up to a complex conjugation; this is reflected by $\boldsymbol{Z}$ being real-valued, which allows the usage of real-valued FFT algorithms. 

\subsection{Odd-Length Case}
Let us now consider the subset of $\boldsymbol{V}_{cso}$ that is odd-length, which we denote by $\boldsymbol{V}_o$. 
Despite the symmetry relation \eqref{eq:v_c_even}, \eqref{eq:S_c_even} does not apply for odd $N$ as \eqref{eq:even_n} does not hold. We can restore \eqref{eq:even_n} by replacing $\boldsymbol{a}_4$ with $\boldsymbol{a}_{4o}[w] = 4w/(N+1), w \in \{0,1, \ldots, N \}$; in other words, we increase the number of sinusoids probed by the DFTs in \eqref{eq:ma_cdfrft} by one such that it is even again. However, we can equivalently append a row of zeros to $\boldsymbol{Z}_o$ before computing column-wise FFTs of length $N+1$.
For this zero-padded $\boldsymbol{Z}_o$, we can now again use \eqref{eq:S_c_even} to halve the number of FFTs required.

\subsection{Computing $\boldsymbol{Z}$ more efficiently} \label{sec:change_of_basis}
Reference \cite{vargas2005multiangle} describes how the symmetries \eqref{eq:v_c_even} within $\boldsymbol{V}_{cso}$ can also be leveraged to halve the number of multiplications necessary to compute \eqref{eq:pre_fft_c}; 
their approach can be easily adapted to $\boldsymbol{V}_{se}$ and its symmetries \eqref{eq:v_s_even}. Efficiency can be further increased by choosing DFT eigenvectors that are sparse and have repeating entries \cite{de2019reduced} or by incorporating ideas from \cite{majorkowska2017low}.
A set of standard DFT eigenvectors where the change of basis into the DFT eigenbasis can be computed in $\mathcal{O}(N \log N)$ \cite{erseghe2006efficient} has been presented in \cite{erseghe2003orthonormal}.

\section{Applications} \label{sec:applications}
While potential applications are manifold (see Sec. \ref{sec:introduction} for some examples), one possible use case of this reduced complexity is the decryption of fractional transform-based cryptographic schemes such as \cite{rawat2012blind}, which utilize the decorrelation between DFRFTs (or related transforms) of different fractional orders.
More concretely, our generalization may reduce the computational complexity of brute-force searching for the secret fractional order, which acts as the key for encrypting data with fractional transforms. Our method might also enable the detection of hidden fractional transform-based watermarks, which can only be detected after computing a fractional transform with a specific secret fractional order. 

Another application is efficient and accurate parameter estimation of LFM chirps; in \cite{peacock2013comparison}, the authors use \cite{vargas2005multiangle} with different CDFT eigenvectors to compare their suitability for LFM chirp parameter estimation. This investigation can now be extended to include standard DFT eigenvectors, possibly resulting in more robust estimates; this is also beneficial for algorithms that use the MA-(C)DFRFT for LFM chirp processing, e.g. \cite{oswald2026fmcw, oswald2025fractional, vargas2005improved, nafchi2021mitigating}. 

\section{Conclusion} \label{sec:conclusion}
In this paper, we reduce the computational complexity of all multiangle fractional $M$-periodic transforms from $\mathcal{O}(N^3)$ to $\mathcal{O}(N^2 \log N)$.
We do this by generalizing the method in \cite{vargas2005multiangle} to arbitrary period-lengths $M$, eigenvalue exponents $\boldsymbol{l}$ and eigenvectors $\boldsymbol{V}$; our generalization now includes the fractional extensions of important transforms such as discrete Fourier, type-I and type-IV discrete sine and cosine, Hadamard and discrete Hartley transforms, among others. 

Moreover, we halve the number of FFTs needed to compute the multiangle fractional discrete Fourier transform (MA-DFRFT) as well as its centered variant, the MA-CDFRFT, by considering the symmetries present in their eigenvectors. The overall complexity to compute an MA-(C)DFRFT is now $\mathcal{O}(N^2)$ for \eqref{eq:pre_fft_c} (ignoring further possible optimizations discussed in Sec. \ref{sec:change_of_basis}) plus $\mathcal{O}((N^2/2) \log N$) for \eqref{eq:ma_cdfrft}; this is a significant reduction compared to a naive implementation with $\mathcal{O}(N^3)$. 

\begingroup
\setlength{\itemsep}{0pt}
\setlength{\parskip}{0pt}
\bibliographystyle{IEEEbib}
\bibliography{refs}
\endgroup

\end{document}